# Spatiotemporal Characterization of VIIRS Night Light

Christopher Small
*Lamont Doherty Earth Observatory*
Columbia University
Palisades, NY USA

csmall@columbia.edu

**Abstract**

The Visible Infrared Imaging Radiometer Suite (VIIRS) Day Night Band (DNB) on board the Suomi NPP satellite now provides almost a decade of daily observations of night light. The temporal frequency of sampling, without the degree of temporal averaging of annual composites, makes it necessary to consider the distinction between apparent temporal changes of night light related to the imaging process and actual changes in the underlying sources of the night light being imaged. The most common approach to night light change detection involves direct attribution of observed changes to the phenomenon of interest. Implicit in this approach is the assumption that other forms of actual and apparent change in the light source are negligible or non-existent. An alternative approach is to characterize the spatiotemporal variability prior to deductive attribution of causation so that the attribution can be made in the context of the full range of spatial and temporal variation. The primary objective of this study is to characterize night light variability over a range of spatial and temporal scales to provide a context for interpretation of night light changes observed on both subannual and interannual time scales. This analysis is based on a combination of temporal moments, spatial correlation and Empirical Orthogonal Function (EOF) analysis. A key result of this study is the pervasive heteroskedasticity of VIIRS monthly mean night light. Specifically, the monotonic decrease of temporal variability with increasing mean brightness. Anthropogenic night light is remarkably stable on subannual time scales. The overall variance partition derived from the eigenvalues of the spatiotemporal covariance matrix are 88%, 2% and 2% for spatial, seasonal and interannual variance (respectively) in the most diverse geographic region on Earth (Eurasia). Heteroskedasticity is a pervasive characteristic of monthly VIIRS composites and is present in all areas for all months of the year, suggesting that much, if not most, of observed month-to-month variability may be related to luminance of otherwise stable sources subjected to multiple aspects of the imaging process varying in time. Given the skewed distribution of all night light arising from radial peripheral dimming of bright sources subject to atmospheric scattering, even aggregate metrics using thresholds must be interpreted in light of the fact that much larger numbers of more variable low luminance pixels may statistically overwhelm smaller numbers of stable higher luminance pixels and cause apparent changes related to the imaging process to be interpreted as actual changes in the light sources.

**Introduction**

The Visible Infrared Imaging Radiometer Suite (VIIRS) Day Night Band (DNB) on board the Suomi NPP satellite now provides almost a decade of daily observations of night light. In comparison to the Defense Meteorological Satellite Program (DMSP) Operational Line Scanner (OLS) night light imagery, VIIRS DNB provides greater dynamic range (14 vs 6 bit), higher spatial resolution (~0.7 vs 5 km), on board calibration and greater sensitivity for low light imaging *(Elvidge et al. 2013)*. All of these features allow VIIRS to detect a much greater diversity of night lights than DMSP-OLS was able to resolve. In addition, VIIRS imagery is available as individual swaths and daily, monthly and annual composites. The greater temporal frequency of sampling, without the degree of temporal averaging of annual composites, makes it necessary to consider the distinction between apparent temporal changes of night light related to the imaging process and actual changes in the underlying sources of the night light being imaged.

Temporal changes in imaged night light arise from a variety of factors related to ambient phenomena (e.g. stray light, lunar cycle, aurora and lightning) (Elvidge et al. 2017; Ji et al. 2018), atmospheric effects *(Fu et al. 2018; Román et al. 2018)*, view geometry *(Li et al. 2019)*, overpass time *(Li et al. 2020)*, background reflectance *(Levin 2017; Levin and Zhang 2017)* and instrument calibration/drift *(Zeng et al. 2018)*, as well as actual changes in terrestrial light sources. Actual changes in light sources also arise from a variety of factors such as disruptions of electricity supply *(Mann, Melaas, and Malik 2016; Kohiyama et al. 2004; Cao, Shao, and Uprety 2013)*, conflict *(Li and Li 2014; Li et al. 2018; Levin, Ali, and Crandall 2017; Li, Chen, and Chen 2013)*, cultural and religious activity *(Roman and Stokes 2015)*, lighted infrastructure development *(Small and Elvidge 2013; Kuechly et al. 2012; Hale et al. 2013; Levin et al. 2014)*, and gas flaring *(Elvidge et al. 2016; Elvidge et al. 2009; Zhang et al. 2015; Coesfeld et al. 2018)*. While the vast majority of research applications focus on changes in actual light sources, the potential contribution of non-source phenomena are often not considered in analyses of night light change. The most common approach to night light change detection involves direct attribution of observed changes to the phenomenon of interest. Implicit in this approach is the assumption that other forms of actual and apparent change in the light source are negligible or non-existent. An alternative approach is to characterize the spatiotemporal variability prior to deductive attribution of causation so that the attribution can be made in the context of the full range of spatial and temporal variation.

The primary objective of this study is to characterize night light variability over a range of spatial and temporal scales to provide a context for interpretation of night light changes observed on both subannual and interannual time scales. Specifically, to introduce a robust methodology for characterization of VIIRS' spatiotemporal variability that can be used to distinguish multiple sources of apparent and actual change in night light. The strategy implemented in this study is based on the combined use of low luminance thresholds and Empirical Orthogonal Function (EOF) analysis of night light time series on both subannual and interannual time scales.

**Data**

The Visible Infrared Imaging Radiometer Suite (VIIRS) sensor was launched on board the NASA-NOAA Suomi satellite in 2011. The day/night band (DNB) of the sensor collects low light imagery in a 3000 km swath at a fixed resolution of 742 m with an equator overpass time of ~1 AM local time. Individual VIIRS acquisitions are often composited to exclude clouds and intermittent sources like fires. In comparison to DMSP, VIIRS provides higher dynamic range, on-board calibration, and multiple optical bands that can be used to distinguish different light sources. More detailed descriptions of the data, products, and applications of VIIRS imagery are given by *(Elvidge et al. 2013)* and *(Miller et al. 2013)*. The VIIRS monthly mean night light composites and cloud free coverages used in this study were produced by the Earth Observation Group at the Colorado School of Mines (https://payneinstitute.mines.edu/eog/). All analyses in this study use the stray-light-corrected monthly mean radiance product. Because VIIRS radiances typically span 4 orders of magnitude, all analyses are performed using $Log_{10}$(radiance)

**Analysis**

This analysis is based on a combination of temporal moment spaces, spatial correlation and Empirical Orthogonal Function (EOF) analysis. Computation of the temporal mean (μ) and temporal standard deviation (σ) of monthly mean radiance and number of cloud free acquisitions per month allows for the combined use of moment composite maps and moment spaces to illustrate the relationship between monthly means and standard deviations. Spatial correlation matrices computed for all pairs of monthly mean radiance images provide a statistical measure of similarity of spatial distributions of night light brightness. Temporal moments and spatial correlations provide complementary aggregate metrics of variability. While these metrics offer the benefit of intuitive interpretability, a more comprehensive depiction of this variability is given by a combined spatiotemporal analysis. Empirical Orthogonal Function analysis, originally developed for statistical weather prediction *(Lorenz 1956)*, is now a standard tool for analysis of spatiotemporal patterns and processes. Overviews of the use of EOF analysis in oceanography and meteorology are given by *(von_Storch and Zwiers 1999; Preisendorffer 1988; Bretherton, Smith, and Wallace 1992)*. The combined use of EOF analysis with temporal feature spaces and temporal mixture models, with application to DMSP night light time series is described by *(Small 2012)*.

EOF analysis uses the principal component transform to represent spatiotemporal patterns as orthogonal modes of variance. Rotating the spatiotemporal coordinate system to align with orthogonal dimensions of uncorrelated variance allows any location-specific pixel time series $P_{xt}$ in an N image time series to be represented as a linear combination of temporal patterns, $F$, and their location-specific components, $C$, as:

$$P_{xt} = \sum_{i=1}^{N} C_{ix} F_{it}$$

(1)

where $C_{ix}$ is the spatial Principal Component (PC) and $F_{it}$ is the corresponding temporal Empirical Orthogonal Function (EOF) and i is the dimension. EOFs are the eigenvectors of the spatiotemporal covariability (either covariance or correlation) matrix that represent uncorrelated temporal patterns of variability within the data. The PCs are the corresponding spatial weights that represent the relative contribution of each temporal EOF to the pixel time series $P_{xt}$ at each location x. The relative contribution of each EOF to the total spatiotemporal variance of the observations is given by the eigenvalues of the covariance matrix. N is the number of discrete dimensions represented by the time series of observations. Principal Components are uncorrelated but not necessarily independent – unless the data are jointly normally distributed. In systems in which the same deterministic processes are manifest at many locations, but stochastic processes are uncorrelated, the variance of the deterministic processes may be represented in the low order PC/EOF dimensions while the stochastic variance may be relegated to the higher order dimensions *(Preisendorfer 1988)*. If a clear distinction can be made between a small number of physically meaningful EOFs (or PCs) distinct from a continuum of uninterpretable EOFs (or PCs), this can provide a statistical basis for attribution of deterministic and stochastic components of an image time series. However, the transformation is purely statistical so there is no guarantee that the attribution will be physically meaningful, or even exist at all.

**Results**

Temporal moment maps illustrate two of the most pervasive sources of subannual variability in VIIRS night light. Figure 1a shows the latitude-dependent variability in solar illumination (or lack thereof) for Europe, southwest Asia and north Africa (henceforth Eurasia) and Figure 1b the effect of summer monsoon cloud cover for southeastern Asia. In both images, the red channel represents the temporal mean of 24 monthly radiance composites for 2019+2020 while the cyan channel represents the temporal standard deviation. The inset maps show the corresponding quantities for the monthly number of cloud free observations contributing to each monthly composite. Both maps for both regions are strongly bimodal, indicating strong regional distinctions between areas with low mean brightness and high temporal variability and areas with high mean brightness and low temporal variability. Areas with high mean brightness and high variability are limited to anthropogenic night lights associated with human settlements and gas flaring. Both high latitude and monsoon summer gaps in coverage result in at least one month without sufficient data to detect even bright lights. Examples are shown in the inset in Figure 1a. Offshore, lights from fishing also results in high variability and low mean brightness as a result of fleets' mobility. In both Eurasia and southeastern Asia the temporal moment spaces (inset) show the bimodal distributions explicitly as two nearly orthogonal limbs corresponding to the aforementioned geographic partitions. The bright+variable night lights appear as diagonal spurs extending away from the high variability limbs of the distributions. The lower, horizontal limb of each distribution represents the more stable night lights corresponding to human settlements and other lighted development. Note that in both distributions, the range of standard deviation diminishes monotonically with increasing mean brightness.

Spatial correlation matrices of VIIRS monthly mean radiance composites for 2019+2020 show correlations between 0.7 and 0.9 for all pairs of months. Tri-temporal color composites of mean radiance for the months with the three lowest spatial correlations are shown in Figure 2. Areas with similar brightness in all three months appear shades of gray. Color implies change. The upper composite shows the anomalously low brightness for large areas of Russia, the Alps and Anatolian plateau in December 2020 as shades of red/orange and a narrow high latitude zone of anomalously high brightness in May 2020 as green along the top of the panel. The lower panel illustrates the effect of seasonal differences. This composite shows the periurban peripheries of lighted developments on the Russian steppe, Alps and Anatolian plateau with higher brightness in January 2020, consistent with the presence of high albedo snow in winter and low albedo vegetation in summer. This is consistent with the seasonality observed by *(Levin 2017)* and *(Levin and Zhang 2017)*. As with the moment spaces in Figure 1, the inset scatterplots in both panels show strong heteroskedasticity with much greater dispersion at lower brightness levels, diminishing monotonically as brightness increases. It is also noteworthy that the considerable skewness magnitude and polarity of the lower tails of all these bivariate distributions varies among all pairs while the upper tail of each distribution remains symmetric about the 1:1 line.

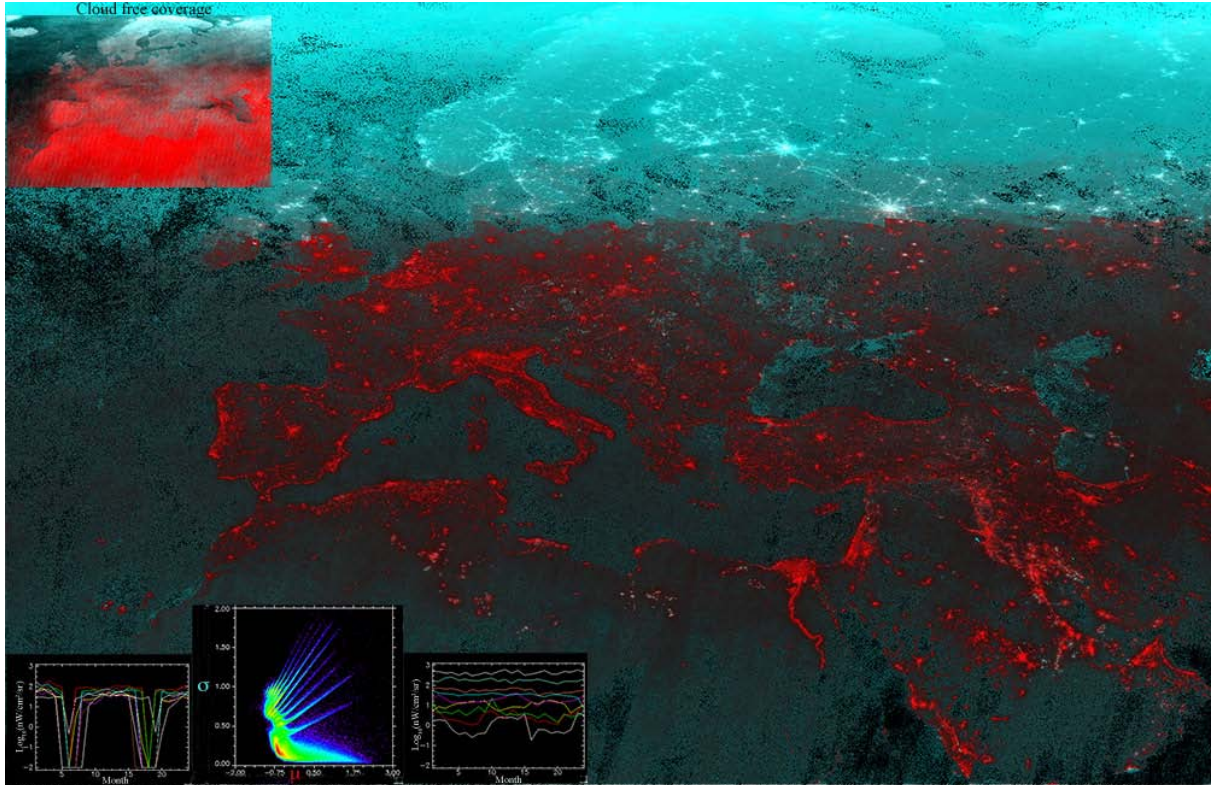

Figure 1a Spatiotemporal moments for $Log_{10}$ VIIRS monthly radiance for 2019+2020. Mean ($\mu$) monthly $Log_{10}$ radiance (red) is consistently high for lighted development and relatively low elsewhere, while the standard deviation ($\sigma$) of monthly $Log_{10}$ radiance (cyan) is low for lighted development and relatively high elsewhere. Lighted development at higher latitudes has high $\mu$ and $\sigma$ so appears white. The $\mu$ vs $\sigma$ distribution (inset scatterplot) shows spurs of rapidly increasing $\sigma$ with $\mu$ for high latitude coverage gaps in summer months. Individual spurs of increasing $\mu$ and $\sigma$ on upper scatterplot correspond to brighter sources with different summer gap lengths (inset left). Monotonically decreasing range of $\sigma$ for $\mu > 1$ nW/cm²/sr on the lowest limb of the distribution results from increasing temporal stability (decreasing $\sigma$) with increasing brightness for sources south of the summer gap latitudes (inset right). Gas flares at all latitudes are both bright and variable (white). Inset map of $\mu$ and $\sigma$ for cloud free acquisitions per month for 2019+2020.(top) shows a similar pattern as a result of cloud cover and high latitude coverage gaps.

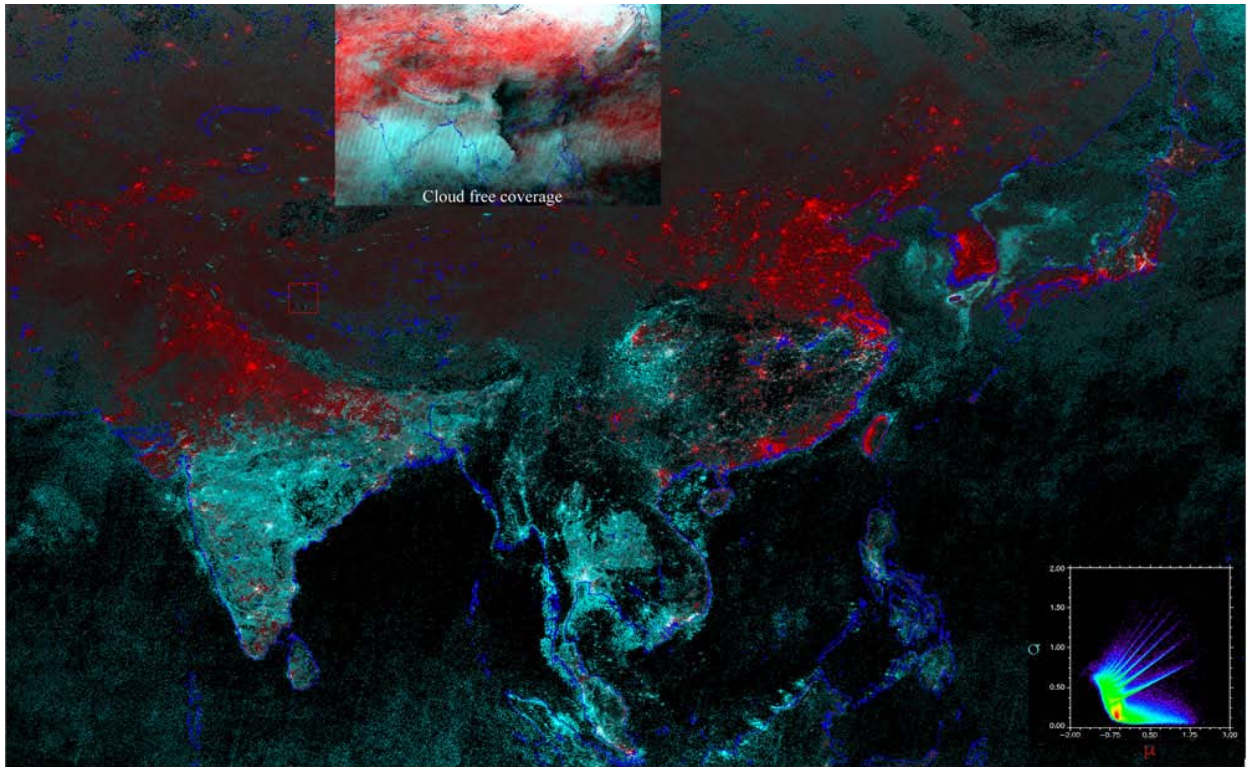

Figure 1b Spatiotemporal moments for $Log_{10}$ VIIRS monthly radiance of Asia for 2019+2020. Mean (µ) monthly $Log_{10}$ radiance is consistently high for lighted development and relatively low elsewhere, while the standard deviation (σ) of monthly $Log_{10}$ radiance (cyan) is low for lighted development and relatively high elsewhere. Lighted development in monsoon Asia has high µ and σ so appears white. The µ vs σ distribution (inset scatterplot) shows spurs of rapidly increasing σ with µ for high latitude coverage gaps in summer months. Individual spurs of increasing µ and σ on upper scatterplot correspond to brighter sources with different summer gap lengths (inset left). Monotonically decreasing range of σ for µ > 0 nW/cm²/sr on the lowest limb of the distribution results from increasing temporal stability (decreasing σ) with increasing brightness for sournce south of the summer gap latitudes (inset right). Inset map of µ and σ for cloud free acquisitions per month for 2019+2020.(top) shows a similar pattern as a result of cloud cover and high latitude coverage gaps.

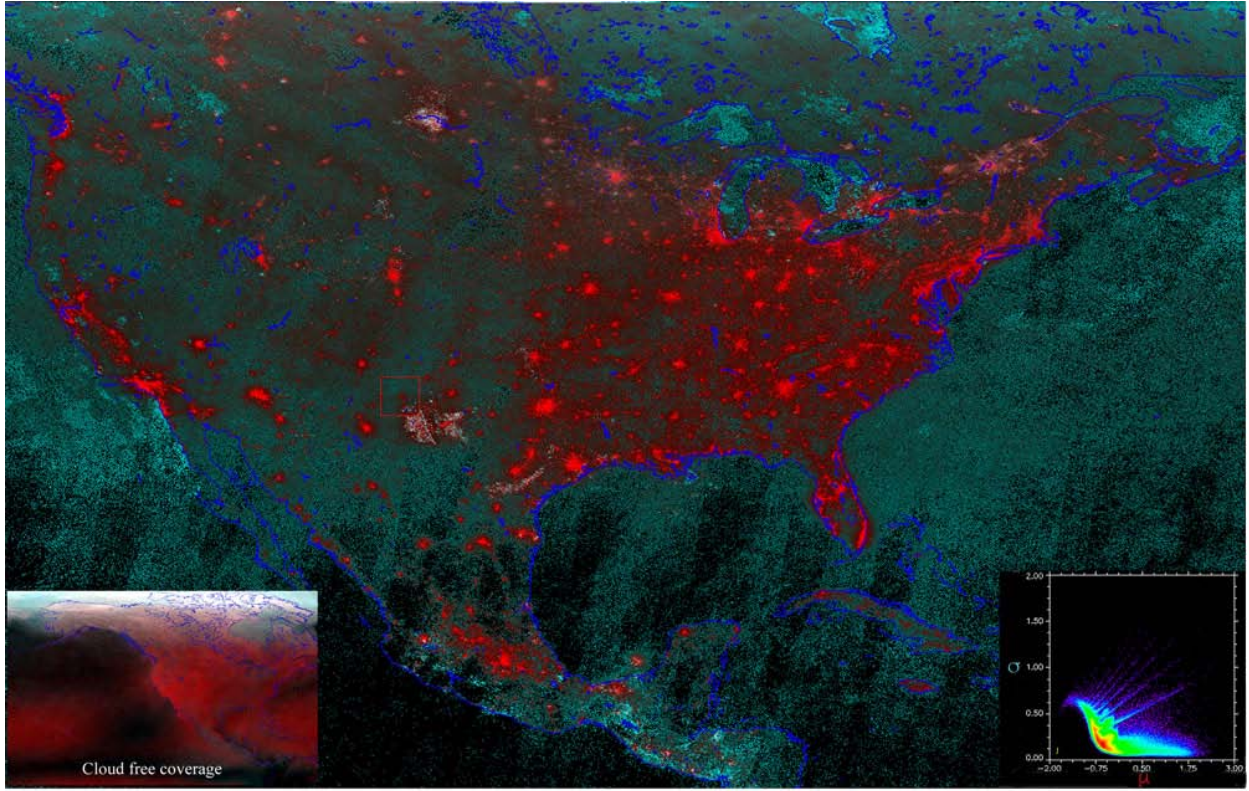

Figure 1 Spatiotemporal moments for $Log_{10}$ VIIRS monthly radiance for 2019+2020. Mean ($\mu$) monthly $Log_{10}$ radiance (red) is consistently high for lighted development and relatively low elsewhere, while the standard deviation ($\sigma$) of monthly $Log_{10}$ radiance (cyan) is low for lighted development and relatively high elsewhere. Lighted development at higher latitudes has high $\mu$ and $\sigma$ so appears white. The $\mu$ vs $\sigma$ distribution (inset scatterplot) shows spurs of rapidly increasing $\sigma$ with $\mu$ for high latitude coverage gaps in summer months. Individual spurs of increasing $\mu$ and $\sigma$ on upper scatterplot correspond to brighter sources with different summer gap lengths (inset left). Monotonically decreasing range of $\sigma$ for $\mu > 0$ nW/cm²/sr on the lowest limb of the distribution results from increasing temporal stability (decreasing $\sigma$) with increasing brightness for sournce south of the summer gap latitudes (inset right). Gas flares at all latitudes are both bright and variable (white). Inset map of $\mu$ and $\sigma$ for cloud free acquisitions per month for 2019+2020.(top) shows a similar pattern as a result of cloud cover and high latitude coverage gaps.
7

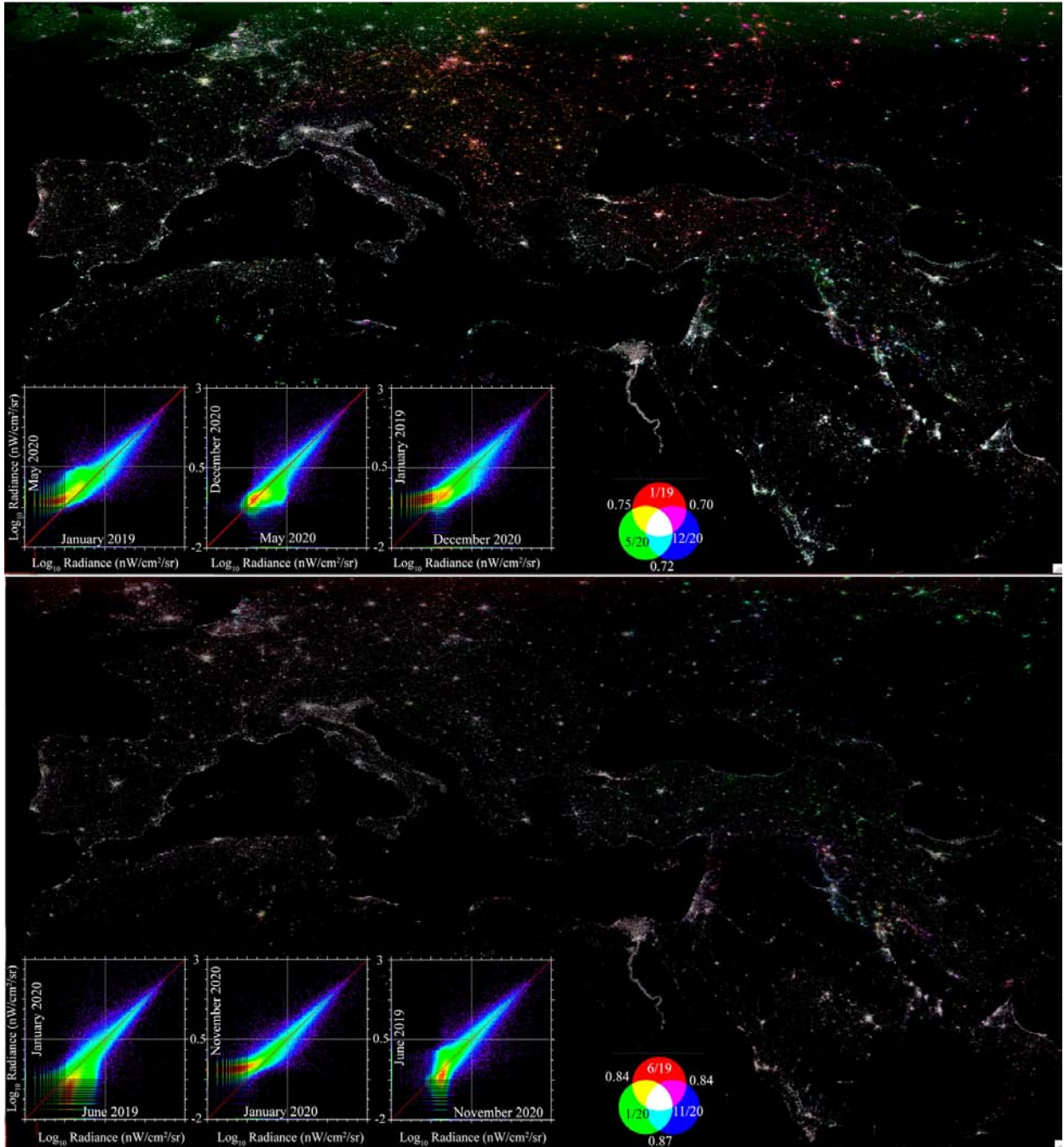

Figure 2 Tri-temporal composites of lowest spatial correlation. Color implies change. May 2020 and December 2020 (top) have much lower correlation to all other months (inset RGB). Aside from these two anomalous dates, summer & winter pairs have lower correlations (bottom) than same season pairs, which are generally > 0.88. Correlations are computed using only pixels > $10^{0.5}$ nW/cm²/sr to exclude effects of background luminance changes. Note geographic contiguity of change areas.

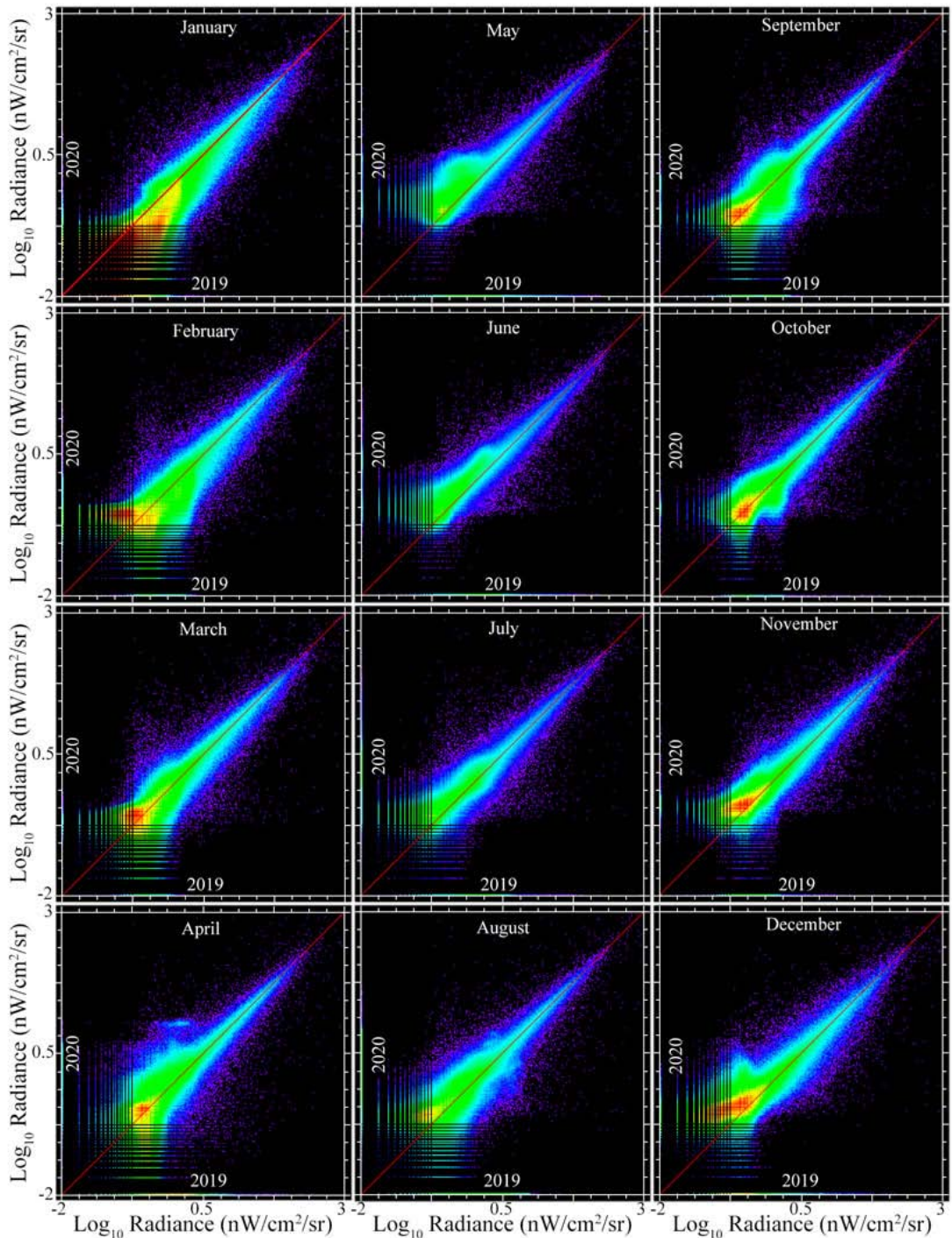

Figure 3 Monthly comparisons of 2019 and 2020 for Eurasia. In all months, dispersion about the 1:1 line diminishes with increasing brightness, with less apparent bias above ~$10^{0.5}$ nW/cm$^2$/sr. At lower brightness levels, there is considerable bias, varying from month to month. Brighter sources show very slight bias suggesting dimming from 2019 to 2020 in January and March 2020, but none apparent in other months.

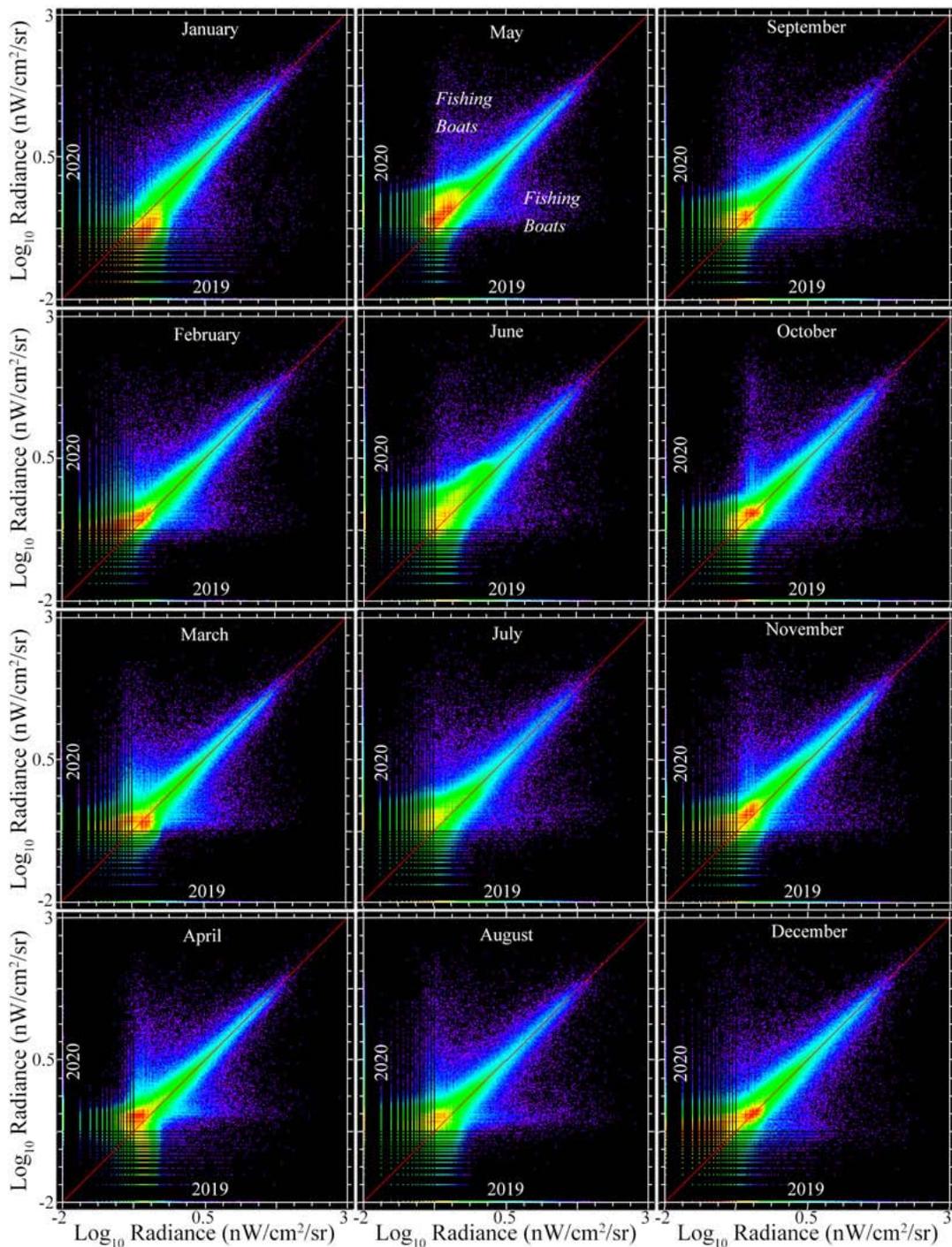

Figure 3b Monthly comparisons of 2019 and 2020. for Asia In all months, dispersion about the 1:1 line diminishes with increasing brightness, with less apparent bias above ~$10^{0.5}$ nW/cm$^2$/sr. At lower levels, there is considerable bias, varying from month to month. Brighter sources show slight bias suggesting dimming from 2019 to 2020 in March and May 2020, but none apparent in other months.

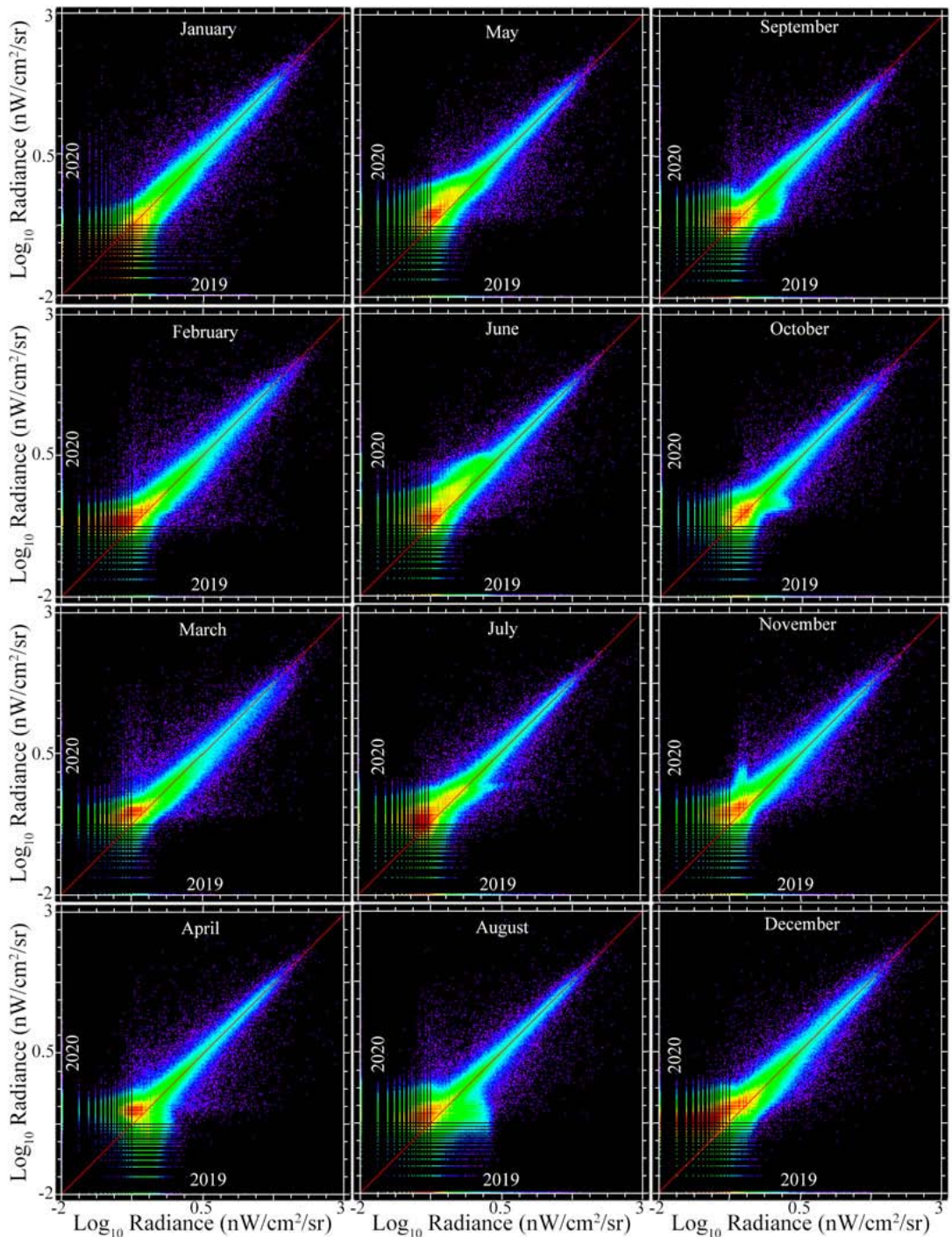

Figure 3c Monthly comparisons of 2019 and 2020 for North America. In all months, dispersion about 1:1 line diminishes with increasing brightness, with less apparent bias above ~$10^{0.5}$ nW/cm²/sr. At lower brightness levels, there is considerable bias, varying from month to month. Brighter sources show slight bias suggesting some dimming in March, September and October, but brightening in April and May.

The seasonal periodicity indicated by the monthly spatial correlation matrix suggests that interannual changes should be most apparent when comparing the same months of different years. Figure 3 shows scatterplots comparing monthly mean radiance for 2019 and 2020 for the Eurasian region shown in Figures 1 and 2. As with the subannual comparisons in Figure 2, the interannual bivariate distributions are all strongly heteroskedastic with skewed lower tails and symmetric upper tails. There is no obvious consistency in the month to month skewness, while the upper tails are consistently symmetric for radiances > $10^{0.5}$ nW/cm$^2$/sr.

The spatial principal components (PCs) of the 24 monthly mean radiance composites from 2019+2020 show the spatial distribution of the dominant modes of spatiotemporal variability for Eurasia driven by the contrast between large numbers of low luminance pixels and much smaller numbers of much brighter pixels (Figure 4). Specifically, a strong contrast between the bright stable lights associated with urban areas and the high variability of background luminance in non-lighted areas. Again the strong contrast between higher latitude and mountainous areas where background reflectance changes seasonally, and lower latitude deserts with negligible seasonality is apparent. This contrast is reflected in the bimodal structure of the temporal feature space of PCs 1 and 2. The structure of the feature space is similar to that of the moment space as the PC1 represents overall brightness varying in space while PC2 represents the dominant mode of temporal variability.

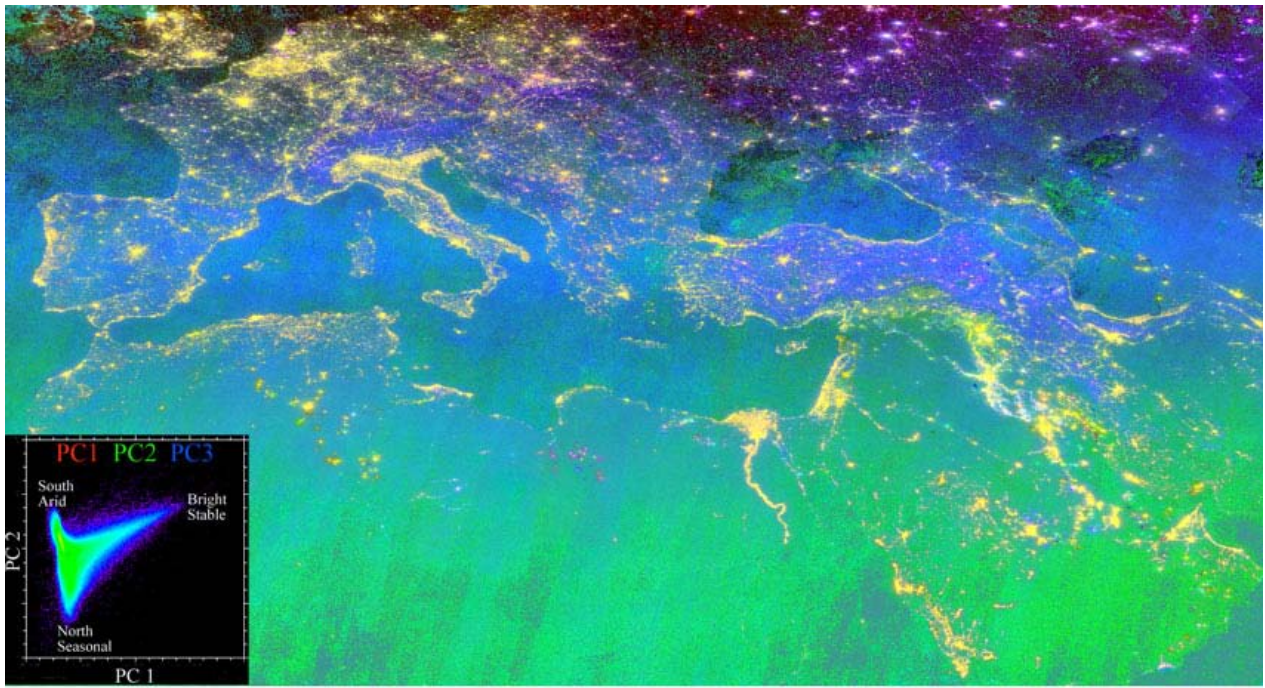

Figure 4 Principal components of the 2019+2020 monthly time series. PC1, PC2 and PC3 account for 48%, 26% and 6% of spatiotemporal variance (respectively). The temporal feature space (inset) shows PC1 corresponding to overall brightness and temporal stability as PC2 represents latitudinal variations in background luminance. PC3 also shows seasonal variability consistent with persistent winter snow cover in mountainous and steppe regions.

In order to suppress the effects of variance heteroskedasticity on the PC rotation, the 24 month image time series was rotated with all pixel time series with radiances less than $10^{0.5}$ nW/cm$^2$/sr masked. Limiting the rotation parameters to those time series with larger radiances effectively focuses the EOF analysis on the variance of anthropogenic light sources. Figure 5 shows a tri-temporal composite of the three low order PCs from the masked time series. As with the other composites, the effect of background reflectance of land cover is clearly apparent – even though it is limited to the larger periurban areas at the scale of the figure. However, it is also clear that there is much greater variability in most of the urban areas, indicated by a greater variety of colors resulting from varying combinations of the low order EOFs. The inset EOFs show a clear separation among spatial variations in overall brightness (PC1 – red), seasonal variations (PC2 – green) and interannual variations (PC3 – blue). Note that all 3 PCs have both positive and negative weights, implying that each corresponding EOF occurs in both polarities. Hence, the upward trend of EOF3 would correspond to decreases in brightness in pixels with negative PC3 values. The overall variance partition derived from the eigenvalues of the covariance matrix are 88%, 2% and 2% for dimensions 1, 2 and 3 respectively. This suggests that almost 90% of the spatiotemporal variance is associated with spatial variations in average brightness while 8% of variance is associated with the remaining higher (> 3) order modes representing stochastic variability and only 2% each for seasonal and interannual variability.

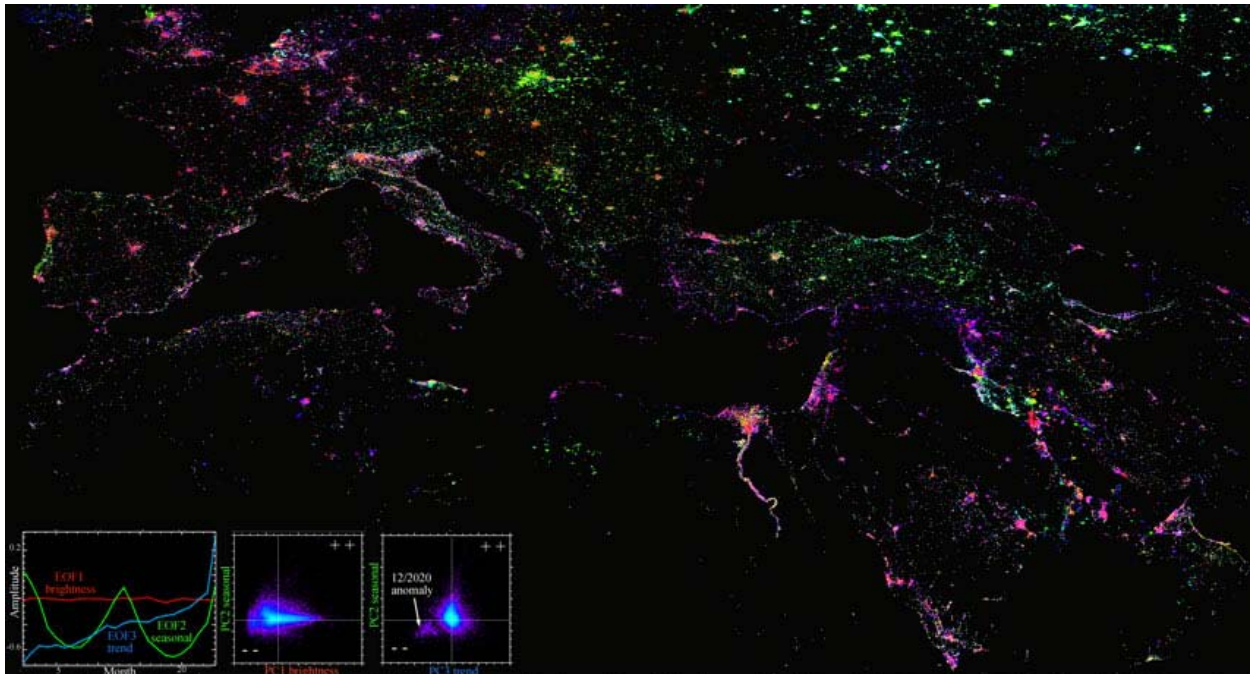

Figure 5 PCs, EOFs and temporal feature space for 2019+2020 monthly time series with luminance threshold applied. PC1, PC2 and PC3 account for 88%, 2% and 2% of spatiotemporal variance (respectively). The > $10^{0.5}$ nW/cm²/sr low luminance threshold eliminates large areas of background luminance so the transformation reflects the spatiotemporal characteristics of the much smaller area of brighter anthropogenic luminance. Bright, temporally stable, urban cores appear red because brightness is modulated by EOF1 & PC1. Magenta and green areas represent variations in the lower luminance periphery where seasonal and interannual changes from residual background luminance approach the contribution of the dimmer anthropogenic light sources. Most of the seasonal (green) areas correspond to mountains or steppe where winter snow and summer vegetation result in higher winter and lower summer albedo. The sign of each PC value determines the polarity of its EOF. The distinct cluster of pixels in quadrant 3 of PC2-PC3 corresponds to parts of eastern Europe and Russia where 12/2020 had anomalously low luminance.

The spectral feature space of the 3 low order dimensions of the low-luminance masked image time series shown in Figure 6 is consistent with both the temporal moment spaces (Figure 1) and the bivariate brightness distributions (Figure 3). As overall brightness (PC1) increases, both seasonal (PC2) and interannual (PC3) variance decreases monotonically. The inset time series taken from the upper edge of the envelope of PC1/PC2 distribution show considerable month to month variability superimposed on the seasonal cycles, consistent with a variance partition between deterministic (PC ≤ 3) and stochastic variance. The PC3/PC2 space is effectively a phase plane in which different combinations of positive and negative PC weights for EOFs 2 and 3 represent different combinations of seasonality and interannual change between 2019 and 2020. Note that the example time series 2 and 4 are not monotonic trends as EOF 3 depicts, but abrupt increases and decreases. This is consistent with the way that abrupt changes are often represented in low order EOFs *(Small 2012)*.

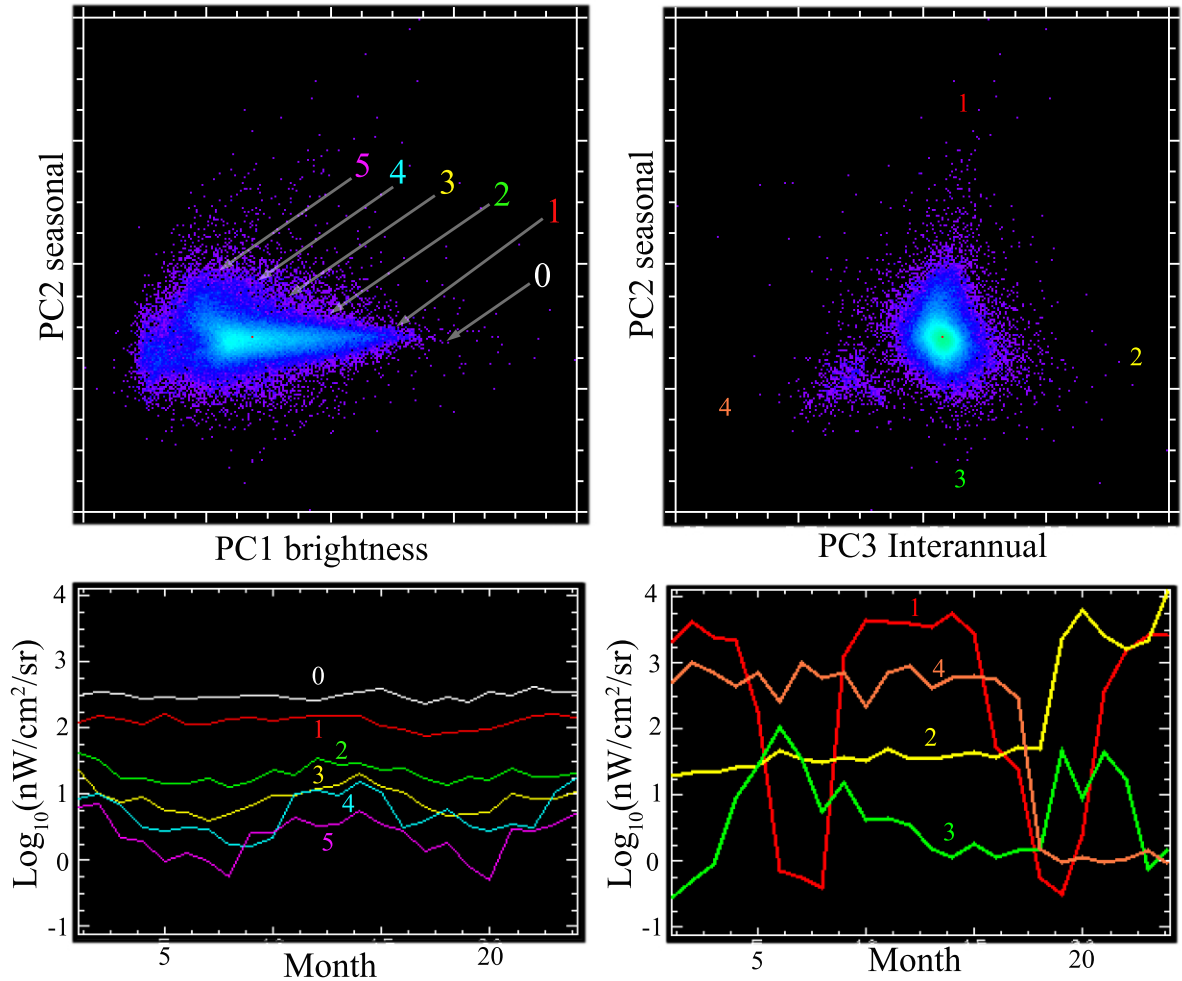

Figure 6 Temporal feature space and example time series for 2019+2020 monthly mean with luminance threshold applied. Seasonal variability diminishes with increasing brightness (left), Brightness changes significantly greater than month to month variability occur, but are exceedingly rare (right).

**Discussion**

*Implications of Subannual Variability*
A key result of this study is the pervasive heteroskedasticity of VIIRS monthly mean night light. Specifically, the monotonic decrease of variability with increasing mean brightness. Bivariate distributions for all month pairs, temporal moment spaces and temporal feature spaces all show this pattern consistently for all years and geographic regions. The three primary sources of geographic variability are latitude-dependent summer gaps in coverage, summer monsoon cloud cover variations and seasonal changes in background reflectance at higher latitudes and elevations where snow is more persistent in winter and vegetation in summer. However, heteroskedasticity is a pervasive characteristic of the monthly composites and is present in all areas for all months of the year, suggesting that much, if not most, of the month-to-month variability may be related to luminance of otherwise stable sources subjected to multiple aspects of the imaging process varying in time. Three of the most obvious that have been documented are viewing geometry *(Li et al. 2019)*, atmospheric opacity related to aerosols *(Fu et al. 2018; Román et al. 2018)* and tropospheric water vapor observed frequently in astronaut photos of cities at night *(Small 2019)*.

EOF analysis quantifies the remarkable stability of brighter sources of night light. Specifically, in urban areas and other types of lighted development. While the two dominant forms of deterministic change (EOF2 and EOF3) account for 2% of variance each, spatial variations in brightness account for almost 90% and stochastic factors account for almost 10%. EOF analysis also illustrates the pervasive influence of background reflectance in dimly lighted periurban and unlighted rural areas – even with a moderate low luminance threshold used to mask background luminance variations.

*Aggregate Metrics of Change*
The heteroskedastic nature of monthly night light composites has important implications for studies that infer actual change in light sources from aggregate metrics like Sum of Lights (SoL) and Number of Pixels (NoP). Even studies that apply low luminance thresholds are subject to variability at the level of the threshold. Given the skewed distribution of all night light arising from radial peripheral dimming of bright sources, even aggregate metrics using thresholds must be interpreted in light of the fact that much larger numbers of more variable low luminance pixels may statistically overwhelm smaller numbers of stable higher luminance pixels and cause apparent changes related to the imaging process to be interpreted as actual changes in the light sources. The aggregation process may conceal the most obvious effects of month-to-month variability, but does so at the risk of misattribution of apparent variability to actual change.

*Spatiotemporal Anomaly Detection*
Because EOF analysis can orthogonalize seasonal and interannual variability as distinct from spatial variability in average brightness, the spatial PCs of these higher dimensions can be used as anomaly detectors to easily identify light sources potentially having undergone actual deterministic change that is statistically distinguishable from both spatial variations in brightness and stochastic temporal variations. Unlike commonly used time series fitting and modeling approaches, EOF analysis is model agnostic and free of the bias inherent in curve fitting. The only assumptions inherent in EOF analysis are that variance corresponds to information and that correlation implies redundancy. In addition, when the combination of variance distribution and EOF interpretability allows for a feasible separation of deterministic and stochastic variance, the resulting partition informs understanding of both. In the case of significant month-to-month or year-to-year variability inconsistent with the nature of stable night light, this variance partition can provide a basis for projection filtering to more clearly isolate the spatial structure of the temporal changes associated with the temporal processes represented by the low order EOFs *(Small 2012; Small and Elvidge 2013)*.